# AI for software engineering: from probable to provable

**Bertrand Meyer**
Bertrand.Meyer@inf.ethz.ch

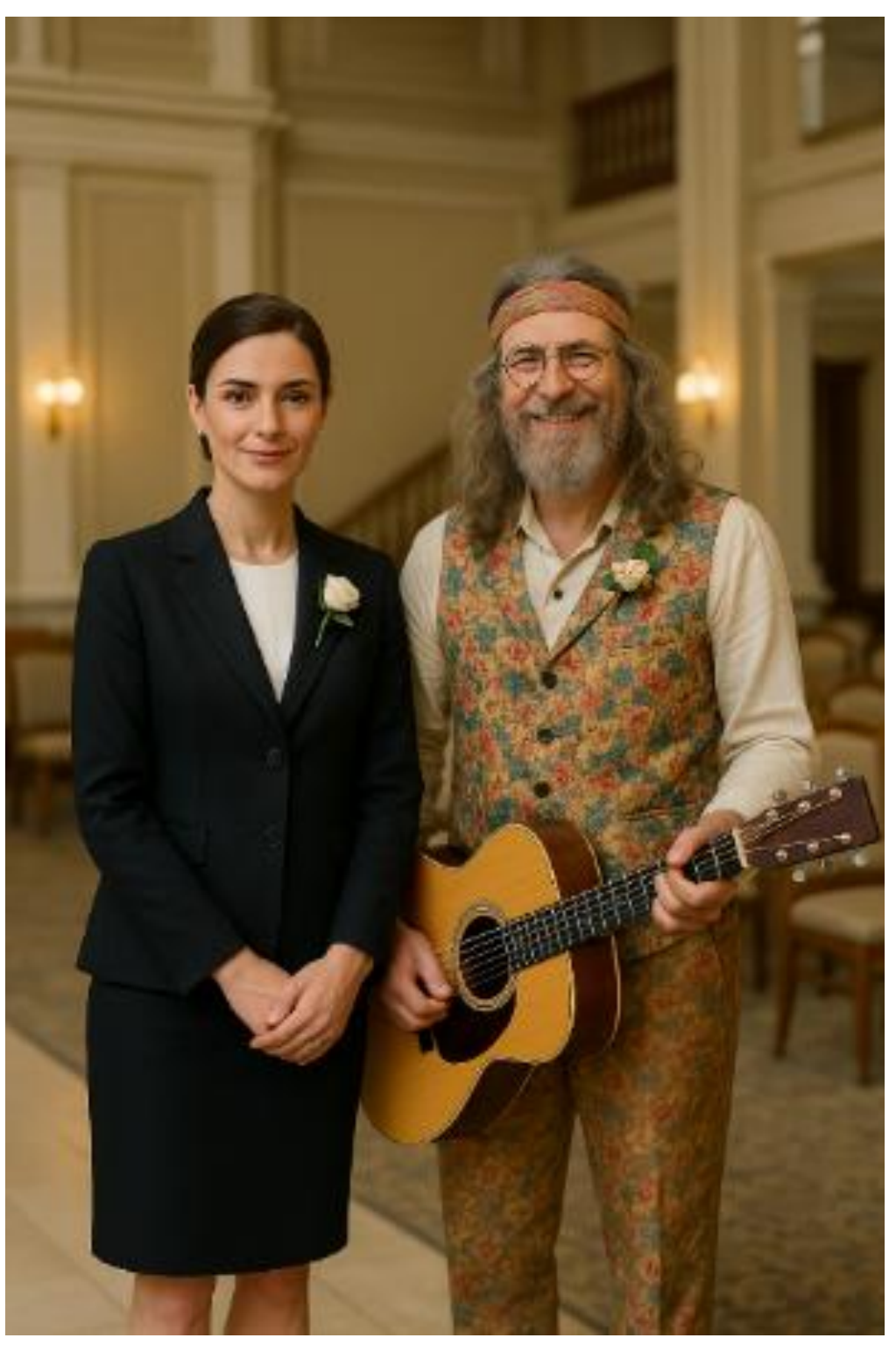

## ABSTRACT

Vibe coding, the much-touted use of AI techniques for programming, faces two overwhelming obstacles: the difficulty of specifying goals ("prompt engineering" is a form of requirements engineering, one of the toughest disciplines of software engineering); and the hallucination phenomenon. Programs are only useful if they are correct or very close to correct.

The solution? Combine the creativity of artificial intelligence with the rigor of formal specification methods and the power of formal program verification, supported by modern proof tools.

Here we go again: no programmers will be needed anymore! AI will generate the code!

If you have been around for a while, you may feel a sense of déjà vu. That same line advertised COBOL in the sixties, 4GLs in the seventies, CASE tools in the eighties, Component-Based Development in the nineties, Model-Driven Architecture in the aughts, low-code/no-code in the tens. Some of these approaches did improve programming, but they did not *replace* programming, let alone programmers. They simply introduced higher levels of abstraction or new tools, sometimes taking advantage of a restricted application domain. Is it the same this time, or do AI and vibe coding upend the game? More generally, can AI and software engineering enter into a successful marriage?

*Warning and spoiler-alert*: even though the following discussion starts out by examining limitations of AI for software construction, do not just expect a critique. Its aim is positive, in *support* of AI-supported software engineering. Its core thesis (here I am really spilling the beans) is that a successful solution requires combining AI with *formal* verification. (End of spoiler.)

## "Prompt engineering" is requirements engineering

Are we about to witness the end of programming? The typical grand pronouncement is something like: "*You will just state what you need: AI will generate the code for you*". The adverb, "*just*", is epic. Ha-ha-ha. Stating what we need ("*just*" what we need) is Requirements Engineering (RE), among the hardest parts of software engineering (I devoted a recent book to it [7]). Anyone who has practiced RE knows that it does not differ that much from programming. It is not exactly the same thing, since it ignores algorithms and the implementation of data structures, but shares many challenges and techniques with programming, particularly concerns of abstraction, structuring, componentization and refinement. Tellingly, many specification languages resemble programming languages, sans the implementation part.

The difficulty of requirements comes in part from the need to specify precise system behavior. Usually, requirements do not specify *all* the details: we are happy to give enough information to enable developers to get started. We also accept slightly *incorrect* requirements: we are happy to assume that programmers will apply common sense. Even if requirements are initially correct, they often become incorrect after a while because the code diverges from them. S*eamless* development, discussed in [7], strives at all times to maintain consistency between all artifacts of software development — requirements, code, designs, tests… — but implies a special software process. If we do use requirements to generate the code through AI, good-enough requirements may no longer be good enough. After all, the code, if correctly generated, will do what we tell the AI tools it should do, rightly or wrongly.

Vibe-coding advocates will dismiss these concerns as naysaying. AI code generation is indeed a reality. Major tech CEOs are on record with statements on how much of their companies' code already comes to life that way, and how much more will in the future. I, too, have been swept away by the impressive results that one gets, *initially*. When trying, for example, to use some existing API: instead of learning it through often haphazard documentation, you can feed your desired scenario to an AI tool and let it figure out how to use the API to realize it.

## Accepting a certain percentage of errors

Problems arise when you move on from experiments, however breathtaking, to real problems. Often, the result somehow *looks* right, but is not.

The problem is that software differs in essential ways from many showcase areas of AI application. In medical analysis, a tool can be OK if it produces the right results most of the time: the alternative — the best human experts — also produces an occasional false negative or positive. If the tool is wrong *less often*, it wins! That is why Modern-AI has already produced a revolution in image analysis. Another revolutionized domain is human-language translation. Until two decades ago, if you came across a document in a language you cannot even decipher (say Korean for me), you were stuck. Today you get a translation, often very good, in seconds. It might still contain a few mistakes, but for a non-speaker it handily beats the alternative (understanding nothing). Even a professional translator can benefit by running the tool to get a rough version then using his expertise to refine the truly difficult parts.

Does this scheme transpose to programming? It can, to some extent. One can imagine a first AI-produced program version, from rough requirements, getting the essentials right (structure and basic algorithms); then programmers correct mistakes. That is what people are attempting to do through vibe coding.

## Software and its correctness

There is a rub: the phenomenon of hallucination. In an early report about experience with LLMs [8], I compared them to a cocky graduate student, smart and widely read, polite, quick to apologize, also sloppy and unreliable. That was harsh but the fact remains that any significant use of current AI tools will produce some wrong advice. At issue are not just catastrophic cases of deleting files instead of moving them, enabling malicious code, or allowing malware by hallucinating package names. These examples, all documented, are not unlike the errors a programmer could make, just like an expert doctor and an AI image analyzer can both misinterpret an X-Ray. The same reasoning applies: if AI can make *fewer* of these mistakes than a programmer would, it is useful. We should not even be turned off by seeing AI faltering on a seemingly natural task, code completion [6]. Such mishaps do not invalidate the larger promises.

More worrying for the long term is what we may call the diabolic case: flawed AI-suggested programs that *seem* right. That is, after all, how Modern-AI works: not by reasoning logically, but by using statistical techniques to produce the most likely answer, based on an enormous training dataset. ("Neurosymbolic" AI [2] seeks to combine probabilistic and logic-based AI, but is still at its inception.)

If an LLM produces a faulty program, you can find and fix the bug, but programming is in practice an iterative process. The diabolic case, called a **hallucination loop** in a recent article [4], arises when a suggestion early on in the iteration sets you off on the wrong path. It looks attractive; you follow it; you see that things do not quite work, but believe you are almost there and ask for improvement; the more you follow the tool's invariably self-confident advice the deeper you dig your hole. In the experiment reported in that article, we presented programmers with buggy programs and offered LLM support. Sometimes the LLM was wrong but looked credible, starting a hallucination loop which goes nowhere. Usually, after a few steps, the programmer gives up. Others have reported similar phenomena [5].

Some might say "these are youthful deficiencies of the technology, they will go away". Being at the start of the Modern-AI era, we should not let superficial limitations derail our judgment (recall how some of the first iPhone reviews picked on minor youthful deficiencies[1]). Hallucinations and particularly hallucination loops seem, however, to be essential, not accidental, to Modern-AI.

Simplistic presentations of AI suggest a continuum in its history, starting in the late 1950s with LISP and the Stanford and MIT AI labs. Such a view ignores that today's technology is something entirely different. Traditional-AI was trying to apply logical reasoning. It failed (for reasons that would deserve another article). The version of AI that has conquered the world — what I have called "modern-AI" — is, as noted, statistics- and probability-based. It computes not the right solution but the most likely solution. An old anecdote captures the transition well: the draft of a large European grant proposal by Tony Hoare, *A Provably Secure Operating System* had a typo in the title, turning the "*v*" into a "*b*"[2]. There we have it.

Showcase AI applications, such as translation and image analysis, benefit tremendously from Modern-AI. For them, a statistically good-enough answer is good enough; the better the approximation (for example, fewer wrong medical diagnoses), the happier we are.

---

[1] Such as the 2007 CNET review: https://www.cnet.com/tech/mobile/original-iphone-review/.

[2] This story sounds like an urban legend but is not. The 1980s ESPRIT project was ProCos (Provably Correct Systems). Prof. Brian Randell, who found and corrected the typo, confirmed the details (private communication, Aug. 2025).

## Software is special

One way software distinguishes itself from other fields is that there are essentially just two kinds of program: a program that works; and a useless program. I know, "*all programs have bugs, that is a fact of life*". Cliché, not always true. Programs "*have bugs*" in the sense of features that do not quite work the way they should, but when they ship they do something useful, and *before they ship* they are typically unusable. Until you have fixed all "critical bugs" you just do not ship, so high are the risks. Few areas of technology have such a correctness exigency on such complex systems.

We can be more precise. An article of a few years ago on "The ABC of software engineering" [9] distinguished three kinds of software system, subject to different criteria. A is for Acute (mission-critical systems, whose malfunctioning may cause catastrophes); C for casual (the other extreme, say a Web site script or the planning of your dance club); and B, in-between, for Business, the mass of enterprise software that has to work but with less momentous consequences than A if it does not.

## When correctness is critical, and when not: the ABC of software engineering

What I wrote above about the almost- all-or-nothing nature of software correctness applies to A and B, covering professional software. The most valid part of the no-more-need-for-programmers prediction arises for category C. It is already common today for people who know little about programming and even less about software engineering to produce a simple application, often in Python, through an AI tool. Entrepreneurs are particularly fond of such vibe coding: you can get a demo or even a Minimum Viable Product for a new project in days — without programmers! That achievement of AI is spectacular and explains the buzz.

Two features characterize many such cases, part of the "C" category: they tolerate quite a bit of deviation from ideal behavior (if your demo crashes for some inputs, you just make sure to avoid them during your presentation to potential investors); and they justify conflating "software development" with just coding, ignoring other software engineering tasks such as requirements, design, verification including testing, maintenance including extensions. For professional software (A, B), these assumptions no longer hold. Reaching a credible level of correctness becomes paramount, and coding typically represents just 10 to 20% of project costs, limiting the overall economic benefit of making that phase faster. Vibe coding does nothing for the rest of software engineering. It may even damage them: see the MIT class experiment where students used an LLM to produce dazzling code but could not explain it, let alone modify it [10].

## Errors can snowball

Another special feature of software is its modular nature, which compounds errors. In his 1970 *Notes on Structured Programming*, Dijkstra remarked that for a system of $N$ modules, each with probability $p$ of being correct, $p$ had better be *extremely* close to 1 if the compounded probability $p^N$ (assuming independence) is to be acceptable. Even with $p$ = 99.9%, for a thousand modules we get about one chance in three (37%) of correctness of the whole. For 5000 modules, the probability is less than 1%!

Unlike Traditional-AI, even very good Modern-AI can only produce a *probably* correct answer; if we multiply probabilities, we cannot come close to guaranteeing the correctness of a realistic system.

## The hippie and the disciplinarian

Correctness is precisely where AI and Software Engineering do not meet. In the marriage picture that illustrates this article, the business suit and stern attitude contrast with the hippie outfit and jolly countenance. The AI partner exudes creativity and empathy. The software engineer is serious, no-nonsense, and interested in just two things: correctness and cost. (Boring but not always bad: as a passenger during landing, you might not wish the flight control system to have been vibe-coded.)

Can the marriage work? After all, software engineering needs creativity along with seriousness. Can we use AI to produce ideas, and software engineering to weed out the bad ideas from the good?

I believe the answer is potentially yes but involves reviving interest in an approach based on mathematical logic and modern tools: *formal specification and verification*. It has actually achieved a respectable level of use in A-type applications (a recent survey [3] covers documented industry deployments), although even there its spread remains limited. In academia, it is not a hot topic (everyone wants to do AI, which is also where the funding goes). Still, it is the inevitable ingredient if AI is to scale up, extend its reach beyond coding, and address professional software engineering, particularly the economically critical "B" (as "A" people will probably always have to do things their own way).

To produce code by specifying — "*just*" specifying — what you want, you need a way to check that the generated implementation, however impressive, complies with it. You cannot hope to do that without having described the specification ("what you want") precisely. Mathematics provides the needed precision. A formal specification can be expressed in a dedicated "specification language" such as B or Alloy, but also in the form of "contracts" embedded in a verification-ready programming language such as Eiffel or Dafny. Then verification tools can determine whether programs meet their specification. That guarantee is a computer-run mathematical proof, which analyzes the program but, unlike tests, does not need to execute it. When it succeeds, it is certain (within the limits of the theory and the tools), unlike the partial hints given by tests. (Tests do retain a role, but that is a topic for another article.)

## Towards a workable process

In case you think the process just sketched — verify all programs we produce — is too rosy, you are right. If it were easy, everyone would be practicing it. Only a minority of projects do.

The main reason appeared at the beginning of this discussion: developing requirements, particularly formal ones, is as hard as developing programs, although in a different way. Anyone who has actually worked to produce formally-verified systems knows a second reason: it is essentially impossible in practice to get either part (specification and implementation) right in one shot. Partly because of conceptual obstacles; partly because of the limitations of verification tools. A realistic (as opposed to textbook) formal verification success requires an iterative process. Specify a little, implement a little (in either order); attempt to verify; hit a property that does not verify; attempt to correct the specification, the implementation or both; repeat. If it sounds familiar, that is because it resembles the practical (as opposed to textbook) way of developing programs, also known as debugging: implement a little, write tests a little (in either order); run tests; hit a bug; attempt to correct it; repeat. The difference is that the iterative process is now static rather than dynamic: we debug not through tests but through proofs.

Here AI is not just part of the problem that leads to seeking such a process integrating development and formal verification; it can also be part of the solution. We can use generative AI to help us produce proposals not only for elements of implementation but also for elements of specification (contracts). AI can, in particular, help in handling one of the thorniest practical tasks of verification: producing *annotations*, such as *loop invariants* and *class invariants*, which many people find daunting. Techniques of machine-learning and mining of large repositories, which have made modern-AI so successful in other fields, seem to offer outstanding potential here, although it largely has to be realized.

To me, that is the process of the future: vibe-contracting complementing vibe-coding. Use the most advanced mechanisms of generative AI, combined with human insight and common sense (the need for which will never go away) to develop specification and implementation hand in hand. At each step, check their consistency with proof tools, such as the one I use, AutoProof for Eiffel (https://autoproof.org) or the Dafny verifier (https://dafny.org), both of which internally rely on the same proof engine, Boogie.

The discipline of combining AI and formal verification is still young, although a number of teams around the globe have shown interesting initial results. A detailed bibliography is beyond the scope of the present contribution but there is already a good survey [1] with pointers to recent articles, and an insightful discussion on neurosymbolic program synthesis [2]. The obstacles are clear, arising from limitations of tools and techniques on both sides: for AI, hallucinations (now a risk for specification in addition to implementation!); for verification, the strictures of today's proof tools, remarkably powerful and sophisticated after decades of progress, but still hard to use.

In spite of the challenges ahead, the combination of AI and formal verification in a tool-supported iterative process seems to be the inevitable path if we want to move from the probable to the provable and make AI and SE live happily ever after — together. Can we avoid a divorce please?

*Acknowledgments* I am grateful to Cesare Pautasso, Olaf Zimmermann, David Lorenz and the CACM referees for important comments.